\documentclass[12pt,twoside]{article}
\usepackage{fleqn,espcrc1}
\usepackage{graphicx}
\usepackage{epsfig}
\usepackage[figuresright]{rotating}

\newcommand{\be}[1]{\begin{equation} \label{(#1)}}
\newcommand{\ee}{\end{equation}}
\newcommand{\ba}[1]{\begin{eqnarray} \label{(#1)}}
\newcommand{\ea}{\end{eqnarray}}

\newcommand{\AmS}{{\protect\the\textfont2
  A\kern-.1667em\lower.5ex\hbox{M}\kern-.125emS}}

\def\be{\begin{equation}}
\def\ee{\end{equation}}
\def\bea{\begin{eqnarray}}
\def\eea{\end{eqnarray}}

\def \gsim {~\mbox{${}^> \hspace*{-9pt} _\sim$}~}

\title{New Physics in the New Millennium with GENIUS: 
	Double Beta Decay, Dark Matter, Solar Neutrinos}

\author{H.V. Klapdor-Kleingrothaus
\address{Max-Planck-Institut f\"ur Kernphysik, 
P.O.Box 10 39 80, D-69029 Heidelberg, Germany\\ 
Spokesman of HEIDELBERG-MOSCOW and GENIUS Collaborations\\
E-mail: klapdor@gustav.mpi-hd.mpg, Home-page: http://mpi-hd.mpg.de.non\_acc/}}

\begin{document}

\maketitle


\begin{abstract}
	Double beta decay is indispensable to solve the question of 
	the neutrino mass matrix together with $\nu$ oscillation experiments. 
	The most sensitive experiment since eight years --- 
	the HEIDELBERG-MOSCOW experiment in Gran-Sasso --- already now, 
	with the experimental limit of 
$\langle m_\nu \rangle < 0.26$~eV excludes degenerate $\nu$ mass scenarios 
	 allowing neutrinos as hot dark matter in the universe for the 
	 small angle MSW solution of the solar neutrino problem. 
	 It probes cosmological models including hot dark matter 
	 already now on the level of future satellite 
	 experiments MAP and PLANCK. 
	 It further probes many topics of beyond Standard Model 
	 physics at the TeV scale. 
	 Future experiments should give access to the multi-TeV 
	 range and complement on many ways the search for new physics 
	 at future colliders like LHC and NLC. 
	 For neutrino physics some of them (GENIUS) will allow to test 
	 almost all neutrino mass scenarios allowed by 
	 the present neutrino oscillation experiments. 
	 At the same time GENIUS will cover a wide range of the parameter 
	 space of predictions of SUSY for neutralinos as cold dark matter. 
	 Further it has the potential to be a real-time detector 
	 for low-energy ($pp$ and $^7$Be) solar neutrinos. 
	 A GENIUS Test Facility has just been funded and will 
	 come into operation by end of 2001.
\end{abstract} 


\section{Introduction}
	Underground physics can complement in many ways the search for 
	New Physics at future colliders such as LHC and NLC and can serve 
	as important bridge between the physics that will be gleaned 
	from future high energy accelerators on the one and, 
	and satellite experiments such 
	as MAP and PLANCK on the other 
\cite{KK60Y,GEN-prop,KK-NOW00,KK-WEIN98,KK-Bey97,KK-SprTracts00,KK-InJModPh98}.

	The first indication for beyond Standard Model (SM) physics 
	indeed has come from underground experiments 
	(neutrino oscillations from Superkamiokande), 
	and this type of physics will play an even large role in the future.

	Concerning neutrino physics, without double beta decay 
	there will be no solution of the nature of the neutrino 
	(Dirac or Majorana particle) and of the structure of the 
	neutrino mass matrix. 
	Only investigation of $\nu$ oscillations 
	{\em and}\ double beta decay together can lead to an 
	absolute mass scale 
\cite{KKPS}--\cite{KKP-ComNucl},\cite{KK-NOON00}.

	Concerning the search for cold dark matter, even a discovery of 
	SUSY by LHC will not have proven that neutralinos form indeed 
	the cold dark matter in the Universe. 
	Direct detection of the latter by underground detectors remains 
	indispensable. 
	Concerning solar neutrino physics, present information on 
	possible $\nu$ oscillations relies on 0.2\% of the solar neutrino flux.
	The total $pp$ neutrino flux has not been measured and also no 
	real-time information is available for the latter.

	The GENIUS project proposed in 1997 
\cite{KK-Bey97,GEN-prop,KK60Y,KK-InJModPh98,KK-J-PhysG98} 
	as the first third generation $\beta\beta$ detector, 
	could attack all of these problems with an unprecedented sensitivity. 
	   In this paper we shall concentrate on the neutrino physics and 
	   dark matter aspects. 
	   The further potential concerning SUSY, compositeness, leptoquarks, 
	   violation of Lorentz invariance and equivalence principle, 
	   etc will only be mentioned briefly and we refer to 
\cite{KK-SprTracts00,KK-InJModPh98,KK60Y,KK-WEIN98,KK-Neutr98}.

	We shall, in section 2, discuss the expectations for the 
	observable of neutrinoless double beta decay, the effective 
	neutrino mass 
$\langle m_\nu\rangle$, from the most recent $\nu$ oscillation experiments, 
	which gives us the required sensitivity for future 
$0\nu\beta\beta$ experiments.
	In section 3 we shall discuss the present status and in section 4 
	the future potential of $0\nu\beta\beta$ experiments.

	It will be shown, that if by exploiting the potential of 
$0\nu\beta\beta$  decay to its ultimate experimental limit, it will be 
        possible to test practically all neutrino mass scenarios allowed 
	by the present neutrino oscillation experiments 
	(except for one, the hierarchical LOW solution).

	In section 5 and 6 we shall outline the potential of GENIUS for 
	dark matter search and for real-time detection of 
	low-energy solar neutrinos.


\section{\boldmath Allowed ranges of 
$\langle m \rangle$ by $\nu$ oscillation experiments}
	 After the recent results from Superkamiokande (e.g. see 
\cite{Suz00,Val01}), the prospects for a positive signal in $0\nu\beta\beta$ 
	  decay have become more promising. 
	  The observable of double beta decay 
$\langle m \rangle =
|\sum U^2_{ei}m^{}_i| = |m^{(1)}_{ee}| 
		      + e^{i\phi_2} |m^{(2)}_{ee}| 
		      + e^{i\phi_3} |m^{(3)}_{ee}|
$
	  with $U^{}_{ei}$ 
	  denoting elements of the neutrino mixing matrix, 
	  $m_i$ neutrino mass eigenstates, and $\phi_i$  relative Majorana 
	  CP phases, can be written in terms of oscillation parameters 
\cite{KKPS,KKPS-01} 
\begin{eqnarray}
\label{1}
|m^{(1)}_{ee}| &=& |U^{}_{e1}|^2 m^{}_1,\\
\label{2}
|m^{(2)}_{ee}| &=& |U^{}_{e2}|^2 \sqrt{\Delta m^2_{21} + m^{2}_1},\\
\label{3}
|m^{(3)}_{ee}| &=& |U^{}_{e3}|^2 \sqrt{\Delta m^2_{32} 
				 + \Delta m^2_{21} + m^{2}_1}.
\end{eqnarray}

	The effective mass $\langle m \rangle$ is related with the 
	half-life for $0\nu\beta\beta$ decay via 
$\left(T^{0\nu}_{1/2}\right)^{-1}\sim \langle m_\nu \rangle^2$, 
        and for the limit on  $T^{0\nu}_{1/2}$
	deducible in an experiment we have 
	$T^{0\nu}_{1/2} \sim a \sqrt{\frac{Mt}{\Delta E B}}$.
	Here $a$ is the isotopical abundance of the $\beta\beta$ emitter;
	$M$ is the active detector mass; 
	$t$ is the measuring time; 
	$\Delta E$ is the energy resolution; 
	$B$ is the background count rate. 
	
	Neutrino oscillation experiments fix or restrict some of the 
	parameters in 
(1)--(3), e.g. in the case of normal hierarchy solar neutrino 
	  experiments yield 
	  $\Delta m^2_{21}$, 
	  $|U_{e1}|^2 = \cos^2\theta_{\odot}$ 
	  and
	  $|U_{e2}|^2 = \sin^2\theta_{\odot}$. 
	  Atmospheric neutrinos fix  
	  $\Delta m^2_{32}$, 
	  and experiments like CHOOZ, looking for $\nu_e$ 
	  disappearance restrict $|U_{e3}|^2$. 
	  The phases $\phi_i$  and the mass of the lightest neutrino, 
	  $m_1$ are free parameters. 
	  The expectations for 
$\langle m \rangle$ 
	  from oscillation experiments in different neutrino mass scenarios 
	  have been carefully analyzed in 
\cite{KKPS,KKPS-01}. In sections 2.1 to 2.3 we give some examples.


\vspace{-0.5cm}
\begin{figure}[ht]
\vspace{9pt}
\centering{
\includegraphics*[scale=0.65]
{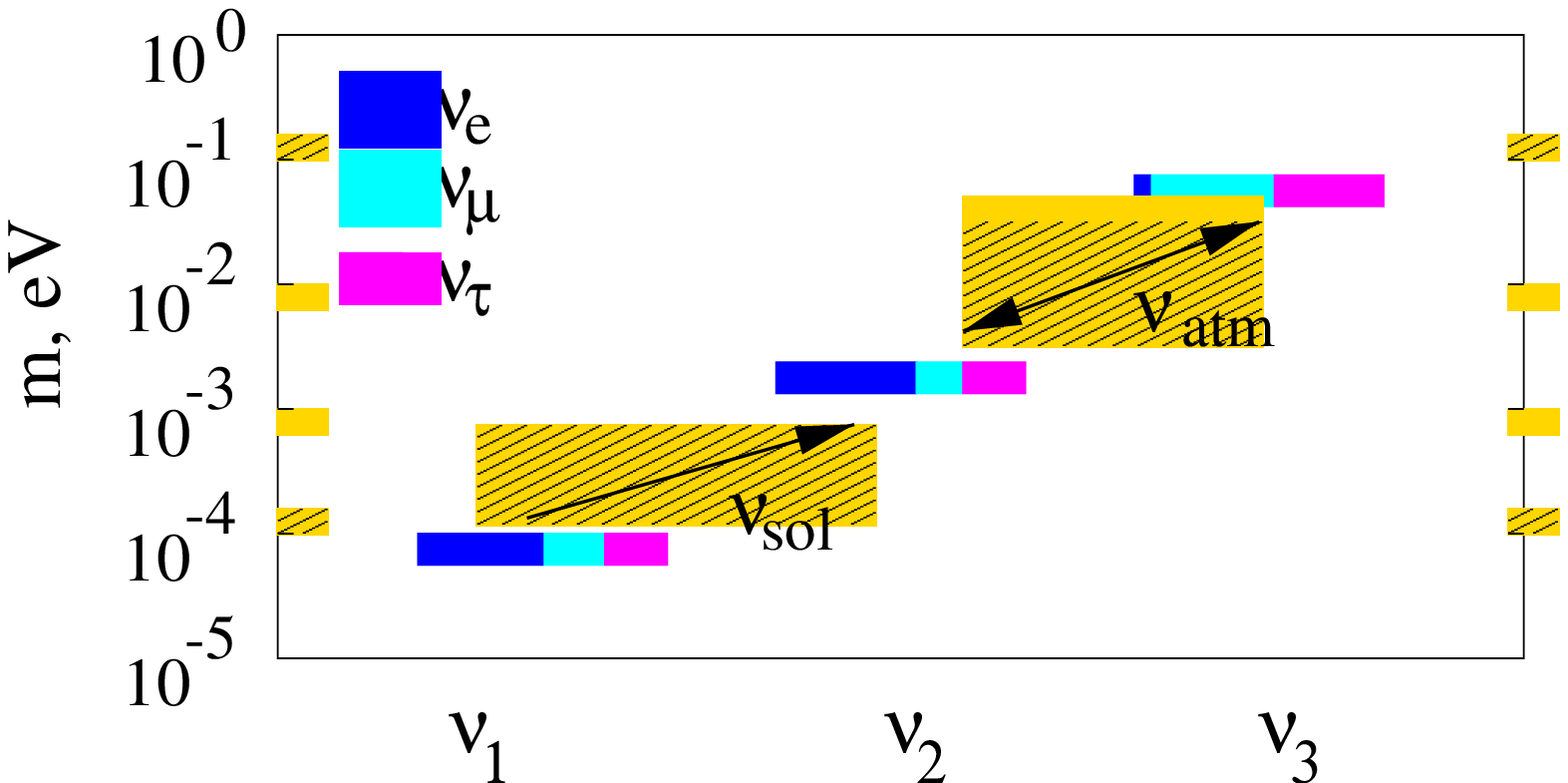}}
\caption[]{
       {\it Neutrino masses and mixings in the scheme with mass hierarchy. 
       Coloured bars correspond to flavor admixtures in the mass 
       eigenstates $\nu_1, \nu_2, \nu_3$. 
       The quantity $\langle m \rangle$
	 is determined by the dark blue bars denoting 
	 the admixture of the electron neutrino $U_{ei}$.}
\label{Hierarchi-NuMass}}
\end{figure}
%


\subsection{\boldmath Hierarchical spectrum $(m_1 \ll m_2 \ll  m_3)$}
         In hierarchical spectra
(Fig.~\ref{Hierarchi-NuMass}), 
	motivated by analogies with the quark sector and the simplest 
	  see-saw models, the main contribution comes from 
	  $m_2$ or $m_3$. 
	  For the large mixing angle (LMA) MSW solution which is favored 
	  at present for the solar neutrino problem (see 
\cite{Suz00}), the contribution of $m_2$ becomes dominant in the expression 
       for $\langle m \rangle$, and  
\begin{equation}
\langle m \rangle \simeq m^{(2)}_{ee} 
	= \frac{\tan^2\theta}{1+\tan^2 \theta}\sqrt{\Delta m^2_{\odot}}.
\end{equation}
	In the region allowed at 90\% C.L. by Superkamiokande according to 
\cite{Val01}, 
	the prediction for $\langle m \rangle$, becomes        
\begin{equation}
\langle m \rangle =(1\div 3) \cdot 10^{-3} {\rm eV}.
\end{equation}
	The prediction extends to 
	$\langle m \rangle = 10^{-2}$ eV in the 99\% C.L. range 
(Fig. ~\ref{Dark2}).



\subsection{\boldmath Inverse Hierarchy $(m_3 \approx m_2 \gg  m_1)$}
\begin{figure}[ht]
\centering{\includegraphics*[scale=0.45]{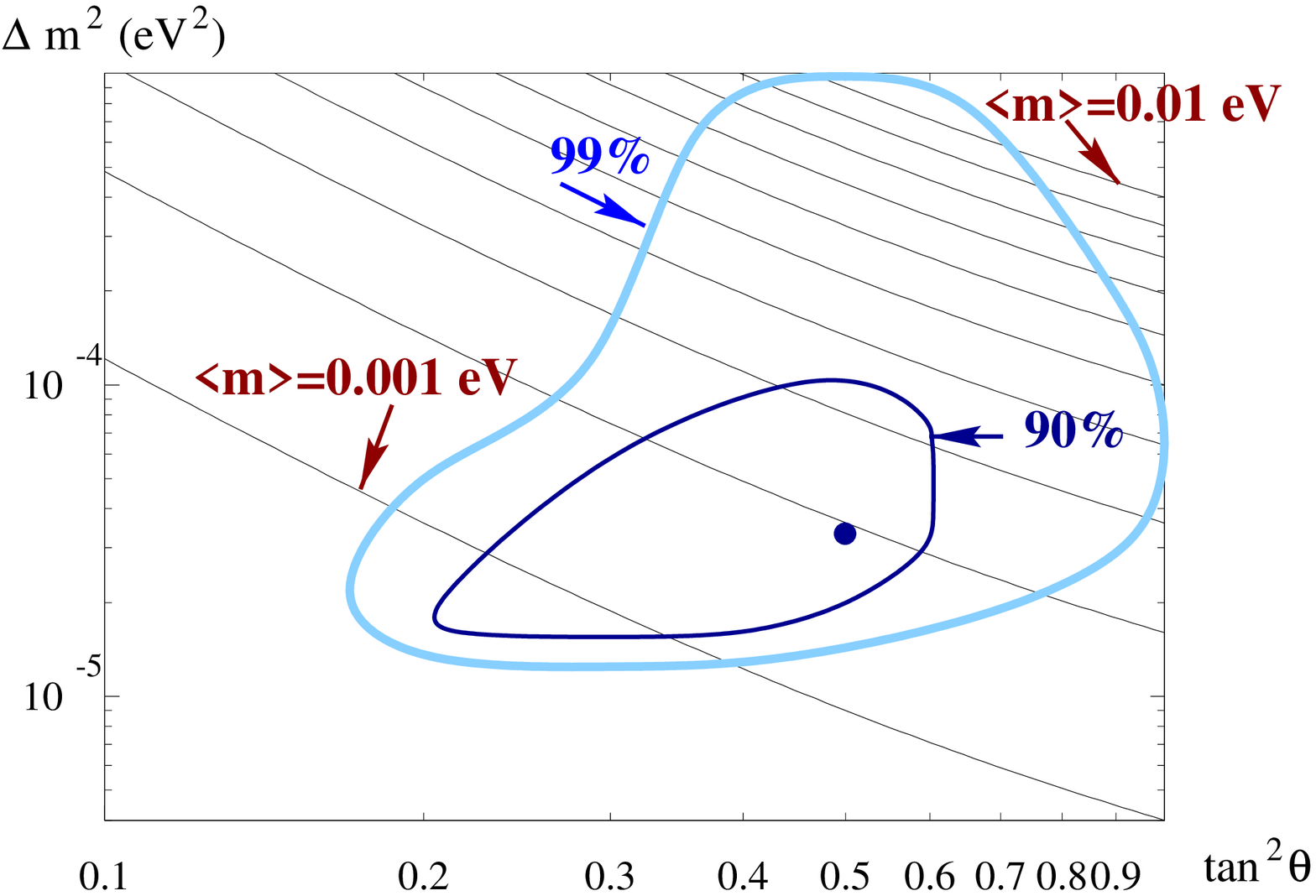}}
\vspace{-0.9cm}
\caption[]{
       {\it Double beta decay observable
$\langle m \rangle$
       and oscillation parameters in the case of the MSW 
       large mixing angle solution of the solar neutrino deficit, 
       where the dominant contribution to $\langle m \rangle$ comes 
       from  the  second  state. 
       Shown are  lines of constant $\langle m \rangle$,  
       the lowest  line corresponding to 
       $\langle m_\nu \rangle = 0.001$~eV, 
       the upper line to 0.01~eV. 
       The inner and outer closed line show the regions allowed 
       by present solar neutrino experiments with 
	90\% C.L. and 99\% C.L., 
       respectively. 
       Double beta decay with sufficient sensitivity could check the 
       LMA MSW solution. 
       Complementary information could be obtained from the search for a 
       day-night effect and spectral distortions in future solar 
       neutrino experiments as well as a disappearance signal in KAMLAND.}
\label{Dark2}
}
\vspace{.5cm}
\centering{
\includegraphics[scale=0.55]
{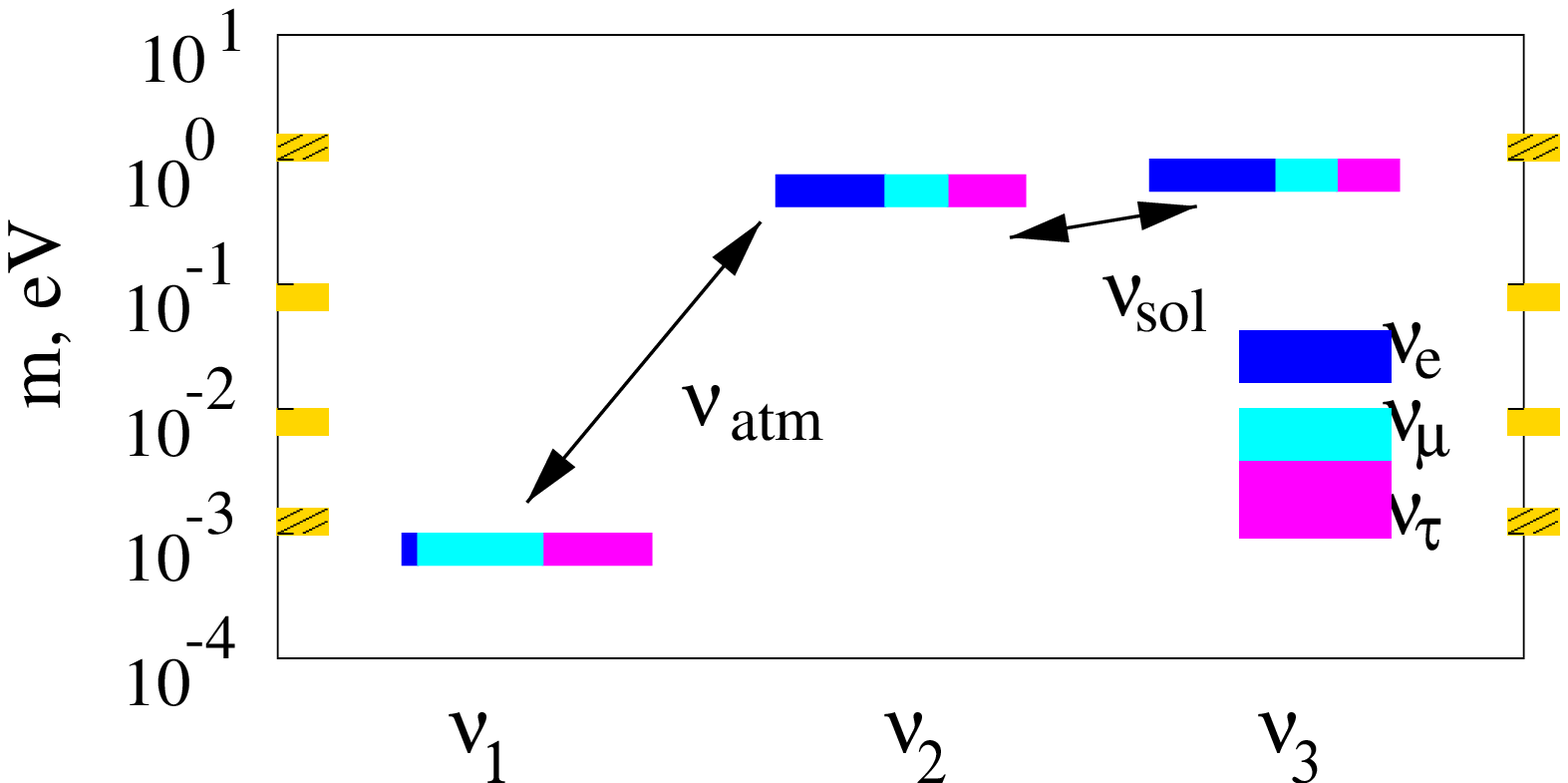}}
\caption[]{
       {\it Neutrino masses and mixing in the inverse hierarchy scenario.}
\label{INVERSE-NuMass}}
\end{figure}

           In inverse hierarchy scenarios 
(Fig.~\ref{INVERSE-NuMass})   
	the heaviest state with mass $m_3$ is mainly the electron 
	   neutrino, its mass being determined by atmospheric neutrinos, 
$m_3 \simeq \sqrt{\Delta m^2_{\rm atm}}$.
	   For the LMA MSW solution one finds 
\cite{KKPS-01}
\begin{equation}
\langle m \rangle 
= (1\div 7) \cdot 10^{-2} {\rm eV}.
\end{equation}



\subsection{\boldmath Degenerate spectrum $(m_1 \simeq m_2 \simeq m_3 
	\gsim	 0.1~$eV)}
        Since the contribution of $m_3$ is strongly restricted by CHOOZ, 
	the main contributions come from $m_1$ and $m_2$, depending on 
	their admixture to the electron flavors, which is determined 
	by the solar neutrino solution. We find 
\cite{KKPS-01}
\begin{equation}
m_{\min} < \langle m \rangle < m_1 \qquad 
\mbox{with} \qquad 
\langle m_{\min}\rangle = 
	(\cos^2\theta_{\odot} -\sin^2\theta_{\odot})\, m^{}_1.
\end{equation}


\clearpage

       	This leads for the LMA solution to 
$\langle m \rangle = (0.25\div 1)\cdot m_1$, 
	 the allowed range corresponding to possible values of 
	 the unknown Majorana CP-phases.

\begin{figure}[htb]
\vspace{9pt}
\centering{
\includegraphics*[scale=0.45, angle=-90]
{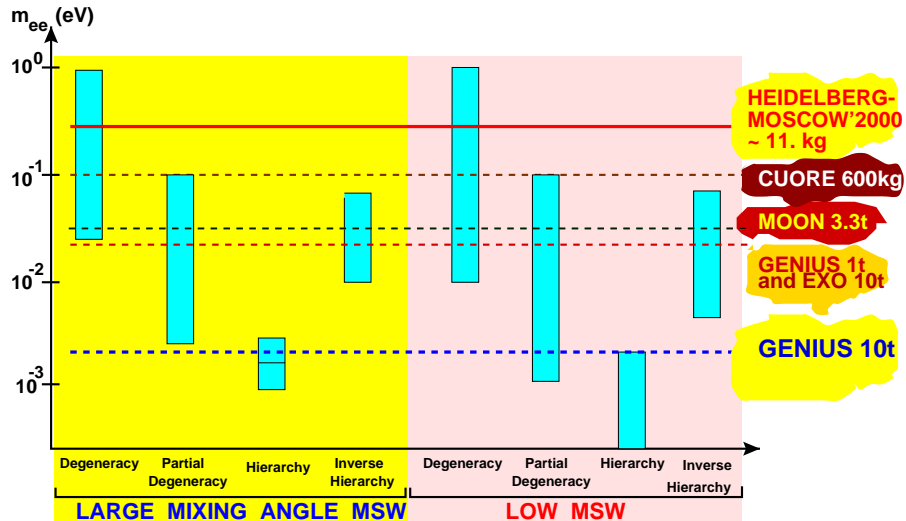}}
\caption[]{
       	{\it Summary of values for $m_{ee} =\langle m \rangle$
	expected from neutrino oscillation experiments 
	(status NEUTRINO2000), in the different schemes discussed in 
	this paper. 
	For a more general analysis see 
\cite{KKPS}. 
	The expectations are compared with the recent neutrino mass 
	limits obtained from the HEIDELBERG-MOSCOW 
\cite{KK01,AnnRepGrSs00}
	, experiment as well as the expected sensitivities for the CUORE 
\cite{CUORE98},  
MOON \cite{Ej00}, 
EXO \cite{EXO00} proposals and the 1 ton and 10 ton proposal of 
      GENIUS 
\cite{KK-Bey97,GEN-prop}.}
\label{Jahr00-Sum-difSchemNeutr}}
\end{figure}

	 After these examples we give a summary of our analysis 
\cite{KKPS,KKPS-01} 
	of the $\langle m \rangle $ allowed by $\nu$ oscillation 
      experiments for neutrino mass models in the presently 
      favored scenarios, 
in Fig.~\ref{Jahr00-Sum-difSchemNeutr}. 
	  The size of the bars corresponds to the uncertainty in 
	  mixing angles and the unknown Majorana CP-phases.


\clearpage


\section{\boldmath Status of $\beta\beta$ Experiments}
\begin{figure}
\centering{
\includegraphics*[scale=0.55, angle=-90 ]
{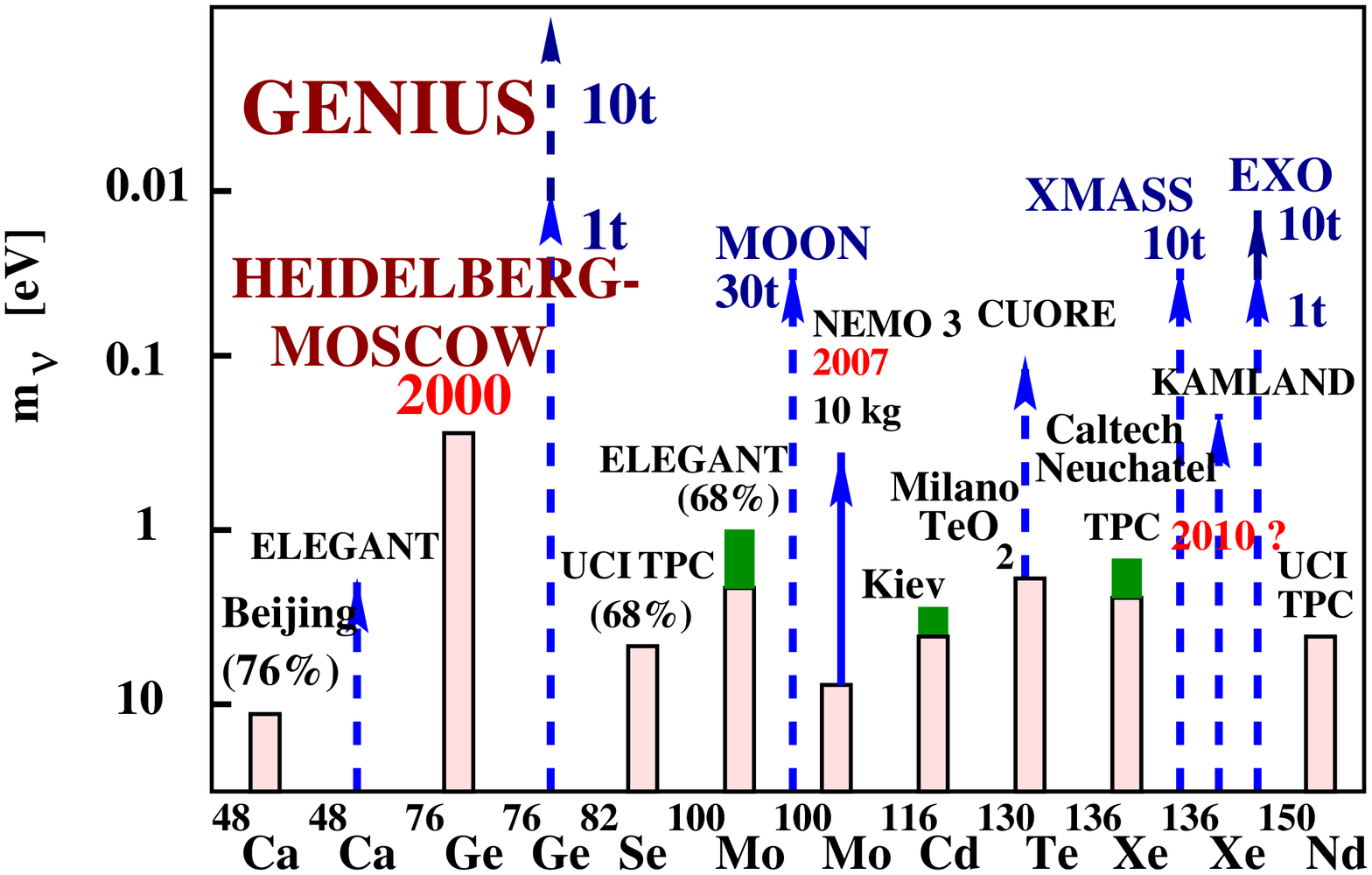}}

\vspace{-.5cm}
\caption[]{
       {\it Present situation, 2000, and expectation for the future, 
       of the most promising $\beta\beta$ experiments. 
       Light parts of the bars: present status; dark parts: 
       expectation for running experiments; solid and dashed lines: 
       experiments under construction or proposed experiments, respectively. 
       For references see} 
\cite{KK60Y,KK-LowNu2,LowNu2}.
\label{Now4-gist-mass}}



\centering{
\includegraphics*[scale=0.45, angle=-90]
{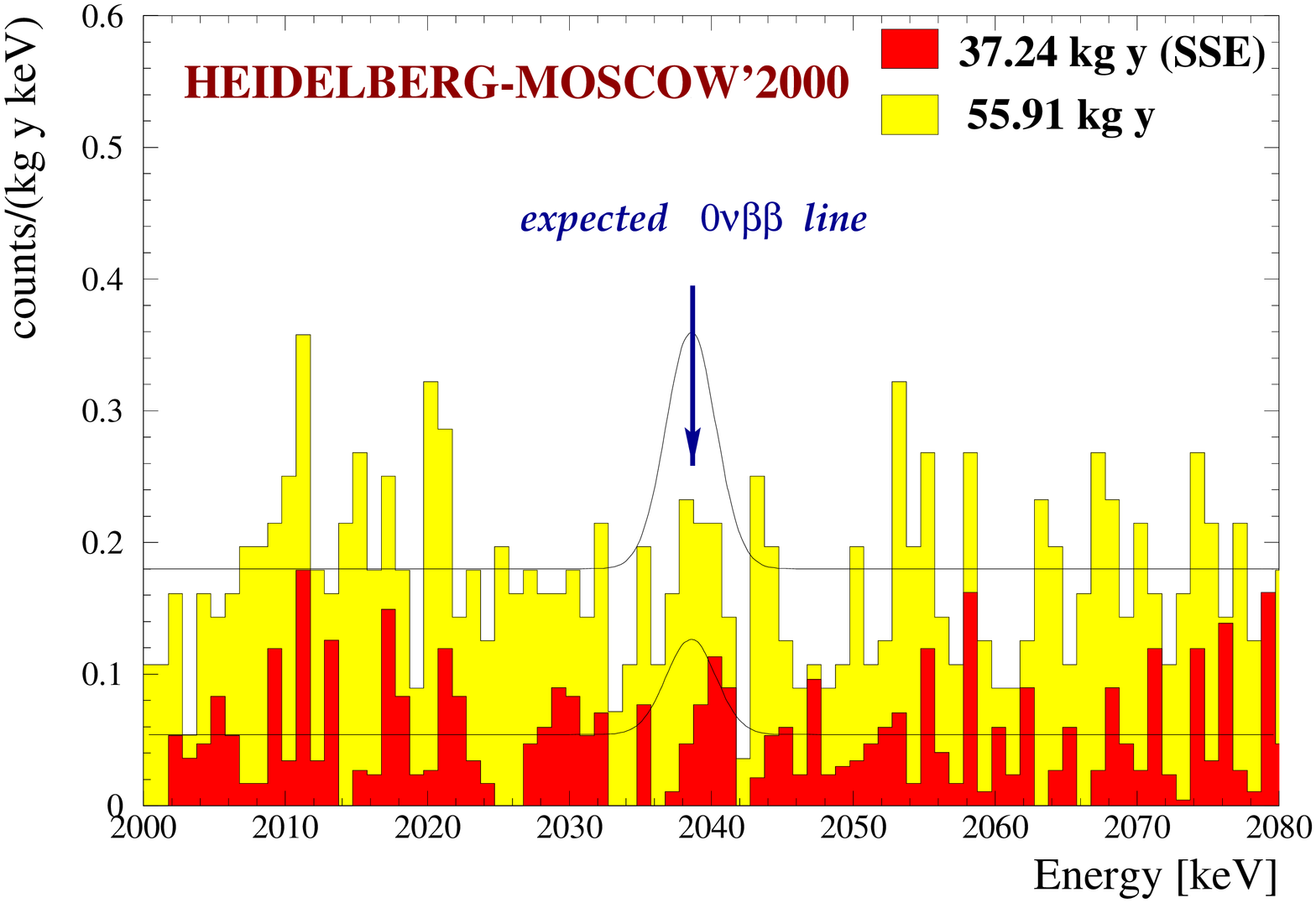}}

\vspace{-.5cm}
\caption[]{
       {\it HEIDELBERG-MOSCOW experiment: energy spectrum in the range 
       between 2000 keV and 2080 keV, where the peak from neutrinoless 
       double beta decay is expected. 
       The open histogram denotes the overall sum spectrum without 
       PSA after 55.9 kg y of measurement (since 1992). 
       The filled histogram corresponds to the SSE data after 37.2 kg y. 
       Shown are also the excluded (90\% C.L.) peak areas from the 
	two spectra.}
\label{Spectr-37-24kgy}}
\end{figure}

	The status of present double beta experiments is shown in 
Fig.~\ref{Now4-gist-mass}	
	and is extensively discussed in 
\cite{KK60Y}.	
	The HEIDELBERG-MOSCOW experiment using the largest source strength 
	of 11 kg of enriched $^{76}$Ge in form of five HP Ge-detectors in 
	the Gran-Sasso underground laboratory 
\cite{KK60Y,KK-StProc00}, 
	yields after a time of 37.2~kg$\cdot$y of measurement
(Fig.~\ref{Spectr-37-24kgy})  a half-life limit of 
\cite{AnnRepGrSs00,AnnRepMPI00} 
$$
T^{o\nu}_{1/2} > 2.1(3.5)\cdot 10^{25}\ {\rm y}, \quad 
	       90\%~(68\%)~{\rm C.L.}
$$
	and a limit for the effective neutrino mass of 
$$
\langle m\rangle 
	< 0.34(0.26)\ {\rm eV}, \quad   
		 90\%~(68\%)~{\rm C.L.}
$$
	This sensitivity just starts to probe some (degenerate) neutrino 
	mass models (see 
Fig.~\ref{Jahr00-Sum-difSchemNeutr}). 
	 In degenerate models from the experimental limit on 
$\langle m\rangle $
	 we can conclude an upper bound on the mass scale of the heaviest 
	 neutrino. 
	 For the LMA solar solution we obtain from 
(7)	 $m_{1,2,3}< 1.1\,$eV implying 
	 $\sum m_i < 3.2\,$eV. 
	 This first number is sharper than what has recently been deduced 
	 from single beta decay of tritium 
	 ($m < 2.2$~eV 
\cite{Weinh-Neu00}),  
	and the second is sharper than the limit of 
	 $\sum m_i < 5.5\,$eV 
	 still compatible with most recent fits of 
	 Cosmic Microwave Background Radiation 
	 and Large Scale Structure data (see, e.g. 
\cite{Teg00}).

\begin{figure}[htb]
\vspace{9pt}
\centering{
\includegraphics*[scale=0.55, angle=-90]
{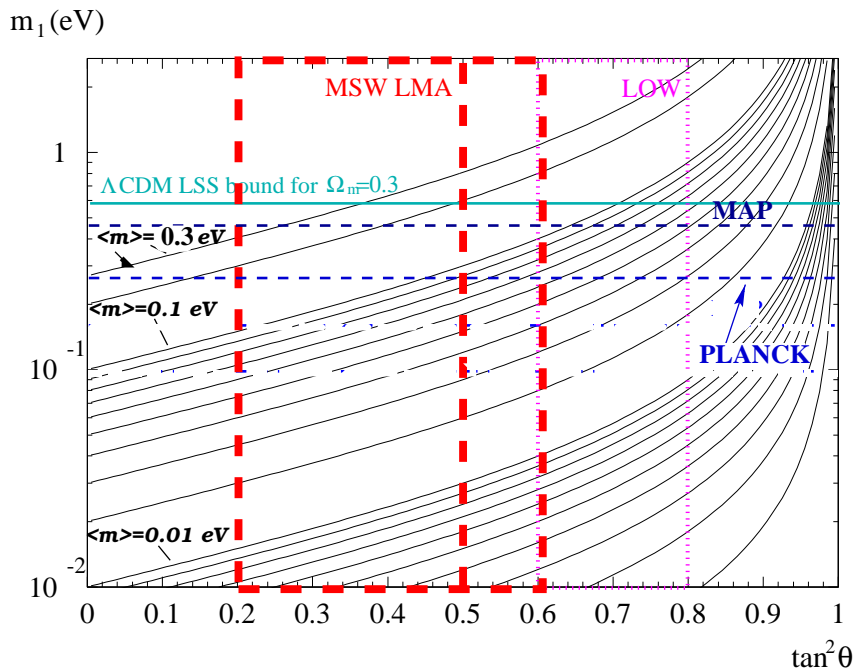}}
\caption[]{
       {\it Double beta decay observable 
$\langle m\rangle$
	 and oscillation parameters: 
	 The case for degenerate neutrinos. 
	 Plotted on the axes are the overall scale of neutrino masses 
	 $m_0$ and mixing $\tan^2\, 2\theta^{}_{12}$. 
	 Also shown is a cosmological bound deduced from a fit of 
	 CMB and large scale structure 
\cite{Lop} 
	and the expected sensitivity of the satellite experiments 
	 MAP and PLANCK. 
	 The present limit from tritium $\beta$ decay of 2.2 eV 
\cite{Weinh-Neu00} 
	would lie near the top of the figure. 
     The range of 
$\langle m\rangle$
	 investigated at present by the HEIDELBERG-MOSCOW experiment is, 
	 in the case of small solar neutrino mixing already in the 
	 range to be explored by MAP and PLANCK 
\cite{Lop}.}
\label{Dark3}}
\end{figure}

	The result has found a large resonance, and it has been shown that 
	it excludes for example the small angle MSW solution of the solar 
	neutrino problem in degenerate scenarios, if neutrinos are 
	considered as hot dark matter in the universe 
\cite{Glash00}--\cite{ElLol99}. 
Figure \ref{Dark3} shows that the present sensitivity 
	 probes cosmological models including hot dark matter already 
	 now on a level of future satellite experiments MAP and PLANCK.


	 The HEIDELBERG-MOSCOW experiment, using the world's largest source 
	 strength, yields now since eight years already the by far sharpest 
	 limits worldwide. 
	 If future searches will show that 
$\langle m\rangle > 0.1$~eV, then the three-$\nu$ mass schemes, 
	 which will survive, are those with $\nu$ mass degeneracy or 
	 4-neutrino schemes with inverse mass hierarchy 
(Fig. \ref{Jahr00-Sum-difSchemNeutr} and 
\cite{KKPS}).

      It has been discussed in detail earlier (see e.g. 
\cite{KK60Y,KK-Bey97,KK-Neutr98,KK-NOW00}
	), that of present generation experiments no one has a 
	   potential to probe 
$\langle m\rangle$
     below the present HEIDELBERG-MOSCOW level (see 
Fig.~\ref{Now4-gist-mass}).

	A second experiment using enriched $^{76}$Ge, IGEX, has stopped 
	operation by end of 1999 
\cite{IGEX00}. 
	This experiment already started in 1992 with 2.1 kg of 
	$^{76}$Ge 
\cite{IGEX93} and operated in 1995 already 8 
     kg of $^{76}$Ge 
\cite{IGEX96}.	In 1999 they published a measuring time of 5.7 kg y 
	(less than one year of full operation) 
\cite{IGEX97,IGEX99}, and in autumn 99 of about 9 kg y 
\cite{IGEX-taup99} 
	 (less than one quarter of the HEIDELBERG-MOSCOW significance) 
	 and an optimistic value for 
$\langle m\rangle$, 
	 using a method criticized.

	 The Milano cryogenic experiment using TeO$_2$ bolometers 
	 improved their values for the 
$\langle m_\nu\rangle$ 
	 from $\beta\beta$ decay of $^{130}$Te, from 5.3 eV in 1994 
\cite{Ales94} to 1.8 eV in 2000 
\cite{Ales00}, and according to 
\cite{CUORE01} to 0.9 eV in early 2001.

	Also CUORICINO (with 45 kg of detectors) scheduled for starting in 
	autumn 2001 
\cite{CUORE01} will hardly reach the HEIDELBERG-MOSCOW limit 
	(see also discussion in 
\cite{Bell00}).

	NEMO-III, originally aiming at a sensitivity of 0.1 eV, 
	reduced their goals recently to $0.3\div0.7$~eV (see 
\cite{NEMO-Neutr00}
	), (which is more consistent with estimates given by 
\cite{Tret95}
	), to be reached in 6 years from starting of running, 
       foreseen for the year 2002.



\section{\boldmath Future of $\beta\beta$ Experiments}
		   To extend the present sensitivity of 
		   $\beta\beta$ experiments below a limit of 0.1 eV, 
		   requires completely new experimental approaches, 
		   as discussed extensively in 
\cite{KK60Y},\cite{KK-Bey97}--\cite{KK-Neutr98}.

	Figure \ref{Jahr00-Sum-difSchemNeutr} 
	shows that an improvement of the sensitivity down to 
$\langle m\rangle\sim 10^{-3}$~eV 
	 is required to probe all neutrino mass scenarios allowed by 
	 present neutrino oscillation experiments 
\cite{KK-Bey97,KKPS}. 
	With this result of $\nu$ oscillation experiments nature seems 
	to be generous to us since such a sensitivity seems to be 
	achievable in future $\beta\beta$ experiment, 
	if this method is exploited to its ultimate limit
\cite{KK60Y,KK-Bey97,GEN-prop}.


\subsection{\boldmath GENIUS, Double Beta Decay and the Light Majorana 
		      Neutrino Mass}
		    
\vspace{.3cm}
	With the era of the HEIDELBERG-MOSCOW experiment 
		    which will remain the most sensitive experiment 
		    for the next years, the time of the small smart 
		    experiments is over.

            The requirements in sensitivity for future experiments to play 
	    a decisive role in the solution of the structure of the 
	    neutrino mass matrix can be read from 
Fig.~\ref{Jahr00-Sum-difSchemNeutr}.

	     To reach the required level of sensitivity $\beta\beta$ 
	     experiments have to become large. 
	     On the other hand source strengths of up to 10 tons of 
	     enriched material touch the world production limits. 
	     At the same time the background has to be reduced by a 
	     factor of 1000 and more compared to that 
	     of the HEIDELBERG-MOSCOW experiment.

Table~1 lists some key numbers for GENIUS, which was the 
	     first proposal for a third generation double beta experiment, 
	     and which may be {\em the only}\ project, which will be able to 
	     test {\em all}\ neutrino mass scenarios, and of the main other 
	     proposals made {\em after}\ the GENIUS proposal. 
	     The potential of some of them is shown also in 
Fig.~\ref{Jahr00-Sum-difSchemNeutr}.	     
	It is seen that not all of these proposals fully cover 
	     the region to be probed. 
	Among them is also the recently presented MAJORANA project 
\cite{MAJOR-WIPP00}, 
	which does not really apply any new strategy for background reduction.

	     The CAMEO project 
\cite{Bell00} 
	will have to work on {\em very}\ long time scales, also 
	     since it has to wait the end of the BOREXINO 
	     solar neutrino experiment.

	     CUORE 
\cite{CUORE-LeptBar98} 
	still has, with the complexity of cryogenic techniques, 
	   still to overcome serious problems of background to enter 
	   into interesting regions of
$\langle m_\nu\rangle$.

	 EXO 
\cite{EXO00} 
	needs still very extensive research and development to probe 
	 the applicability of the proposed detection method.

	 In the GENIUS project a reduction by a factor of more than 1000 
	 down to a background level of 0.1 events/tonne y keV 
	 in the range of $0\nu\beta\beta$ decay is reached by removing all 
	 material close to the detectors, and by using naked Germanium 
	 detectors in a large tank of liquid nitrogen. 
	 It has been shown that the detectors show excellent 
	 performance under such conditions 
\cite{GEN-prop,KK-Bey97}.

	For technical questions and extensive Monte Carlo simulations of 
	the GENIUS project for its application in double beta decay 
	we refer to 
\cite{GEN-prop,KK-J-PhysG98}.



\subsection{GENIUS and Other Beyond Standard Model Physics}
		   GENIUS will allow besides the major step in neutrino 
		   physics described above the access to a broad range 
		   of other beyond SM physics topics in the multi-TeV range. 
	Already now $\beta\beta$ decay probes the TeV scale on which new 
	physics should manifest itself (see, e.g. 
\cite{KK-Bey97,KK-J-PhysG98,KK-LeptBar98}). 
	  Basing to a large extent on the theoretical work of the 
	  Heidelberg group in the last five years, the 
	  HEIDELBERG-MOSCOW experiment yields results for SUSY models 
	  (R-parity breaking, neutrino mass), leptoquarks 
	  (leptoquarks-Higgs coupling), compositeness, right-handed $W$ mass, 
	  nonconservation of Lorentz invariance and 
	  equivalence principle, mass of a heavy left or 
	  righthanded neutrino, competitive to corresponding results 
	  from high-energy accelerators like TEVATRON and HERA. 
	  The potential of GENIUS extends into the multi-TeV region for 
	  these fields and its sensitivity would correspond to that of 
	  LHC or NLC and beyond (for details see 
\cite{KK60Y,KK-J-PhysG98,KK-LeptBar98,KK-SprTracts00}).


\section{GENIUS and Cold Dark Matter Search}
\begin{figure}
\begin{picture}(100,145)
\centering{
\put(100,-170){\includegraphics{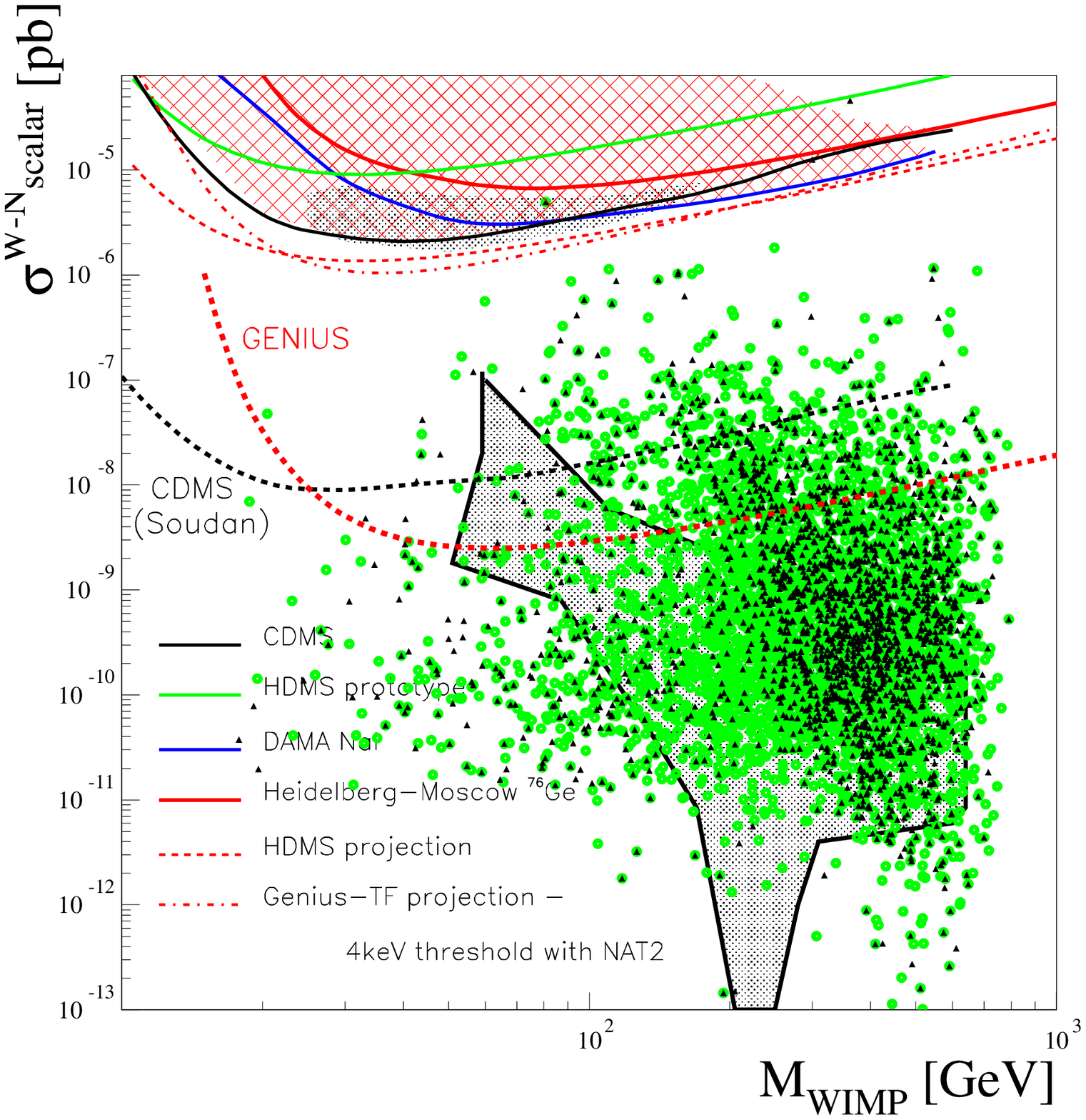}}}
\end{picture} 

\vspace{2.3cm}
\caption[]{
       {\it WIMP-nucleon cross section limits in pb for scalar interactions as 
       function of the WIMP mass in GeV. 
       Shown are contour lines of present experimental limits (solid lines) 
       and of projected experiments (dashed lines). 
       Also shown is the region of evidence published by DAMA. 
       The theoretical expectations are shown by a scatter plot (from 
\cite{BedKK00}) and by grey region (from 
\cite{EllOliv-DM00}). 
	{\em Only}\ GENIUS will be able to probe the shown range 
       also by the signature from seasonal modulations.}
\label{Bedn-Wp2000}}



\vspace{-1.3cm}
\centering{
\includegraphics*[scale=0.50, angle=-90]
{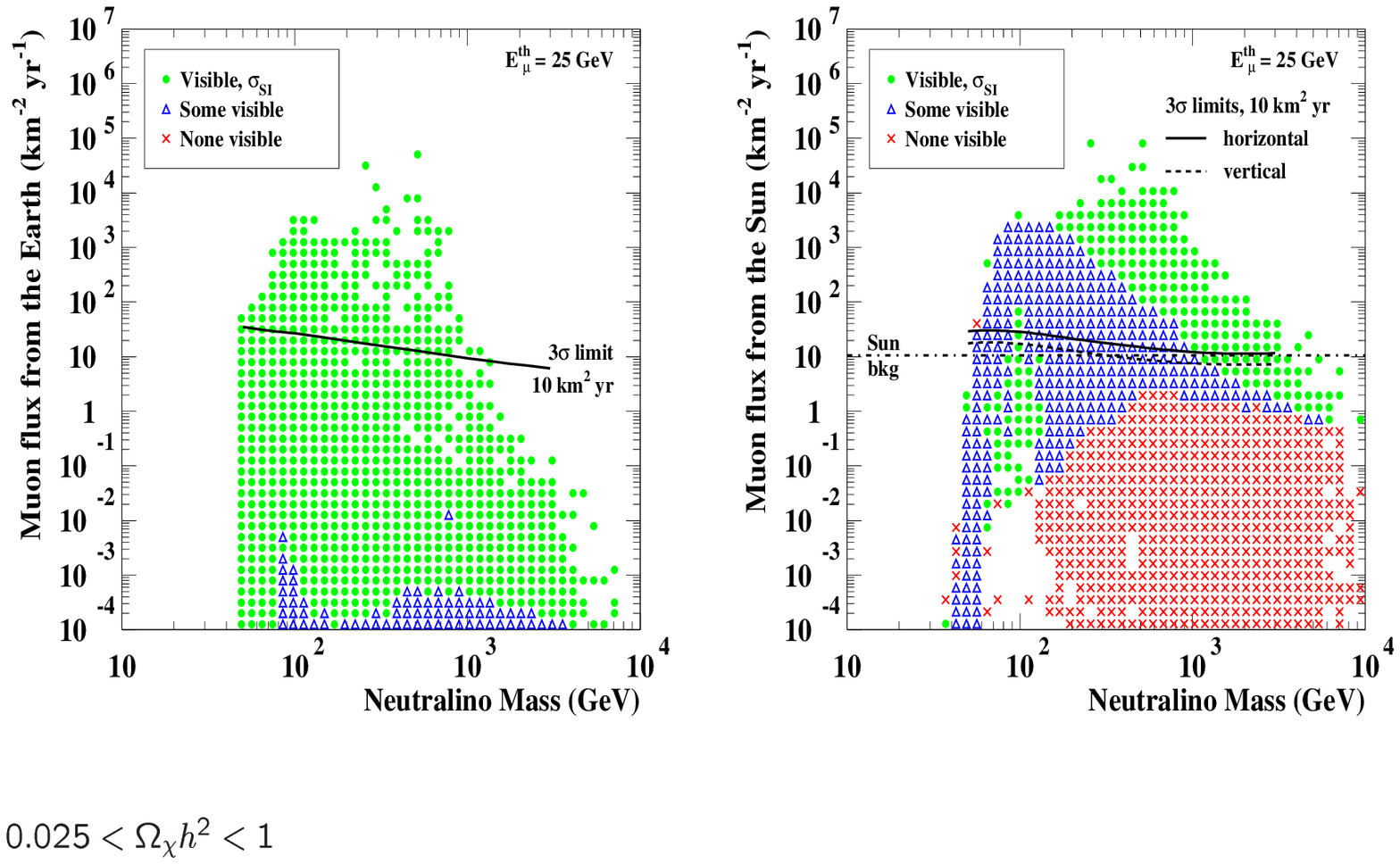}}

\vspace{-2.cm}
\caption[]{
       {\it Sensitivity of future AMANDA 1 km$^3$ for indirect detection 
       of dark matter and of GENIUS for direct detection of dark matter. 
       AMANDA looks for muons from neutrinos produced by neutralino 
       annihilation, and can exclude (probe) the range beyond the solid time. 
       GENIUS looks for nuclear recoils from neutralino scattering. 
       It can probe the light shaded area in the left figure, 
       i.e. is much more sensitive than AMANDA for neutrinos 
       from the Earth, and the upper and middle light, as well as 
       partly the dark areas in the right figure, 
       i.e. is of similar sensitivity as 
       AMANDA for neutrinos from the Sun (from 
\cite{Eds99}).}
\label{Amanda-New}}
\end{figure}

		Already now the HEIDELBERG-MOSCOW experiment is the most 
		sensitive Dark Matter experiment worldwide concerning 
		the raw data 
\cite{KK-PRD63-00,HM98,Ram99,KK-dark00}. 
	GENIUS would already in a first step, with 100 kg of 
		{\it natural} Ge detectors, cover a significant part of the 
		MSSM parameter space for prediction of neutralinos 
		as cold dark matter 
(Fig.~\ref{Bedn-Wp2000}) 
	(see, e.g.
\cite{KKRam98}). For this purpose the background in the energy range 
		$< 100$~keV has to be reduced to 
		$10^{-2}$ events/(kg y eV), 
		which is possible if the detectors are produced 
		and handled on Earth surface under heavy shielding, 
		to reduce the cosmogenic background produced by 
		spallation through cosmic radiation (critical products are 
		tritium, $^{68}$Ge, $^{63}$Ni, ...) to a minimum. 
		For details we refer to 
\cite{GEN-prop,DARK2000}. 
Fig.~\ref{Bedn-Wp2000} 
	shows together with the expected sensitivity of GENIUS, 
     predictions for neutralinos as dark matter by two models, one 
     basing on supergravity 
\cite{EllOliv-DM00}, another starting from more 
     relaxed unification conditions 
\cite{BedKK00}.

	     The sensitivity of GENIUS for Dark Matter corresponds to 
	     that obtainable with a 1 km$^3$ AMANDA detector for 
	     {\it indirect} detection (neutrinos from annihilation 
	     of neutralinos captured at the Sun) (see 
Fig.~\ref{Amanda-New}) \cite{Eds99}. 
	Interestingly both experiments would probe different neutralino 
	compositions: GENIUS mainly gaugino-dominated neutralinos, 
	AMANDA mainly neutralinos with comparable gaugino and 
	Higgsino components (see 
Fig. 38 in 
\cite{Eds99}). 
     It should be stressed that, together with DAMA, GENIUS will be 
     {\em the only}\ future Dark Matter experiment, which would be able to 
     positively identify a dark matter signal by the seasonal 
     modulation signature. 
     This cannot be achieved, for example, by the CDMS experiment.


\begin{figure}
\centering{
\includegraphics*[scale=0.45, angle=-90]
{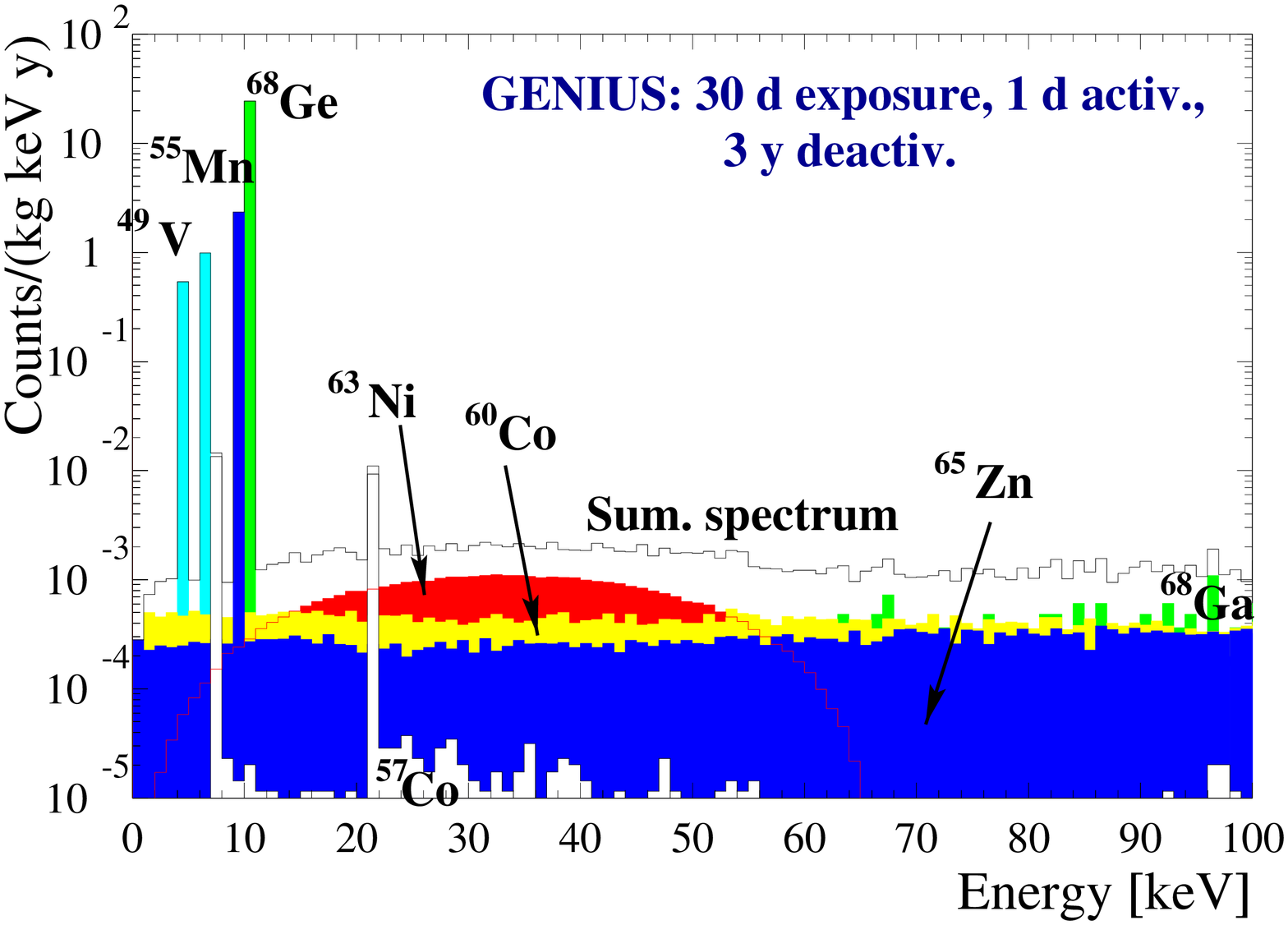}}
\caption{
       {\it Simulated cosmogenic background during detector production. 
       Assumptions: 30 days exposure of material before processing, 
       1 d activation after zone refining, 3 y deactivation 
       underground (neglecting tritium production) (see
 \cite{KK01,KK-LowNu2}).}
\label{Cosmo-1d-3y}}
%
\centering{
\includegraphics*[scale=0.45, angle=-90]
{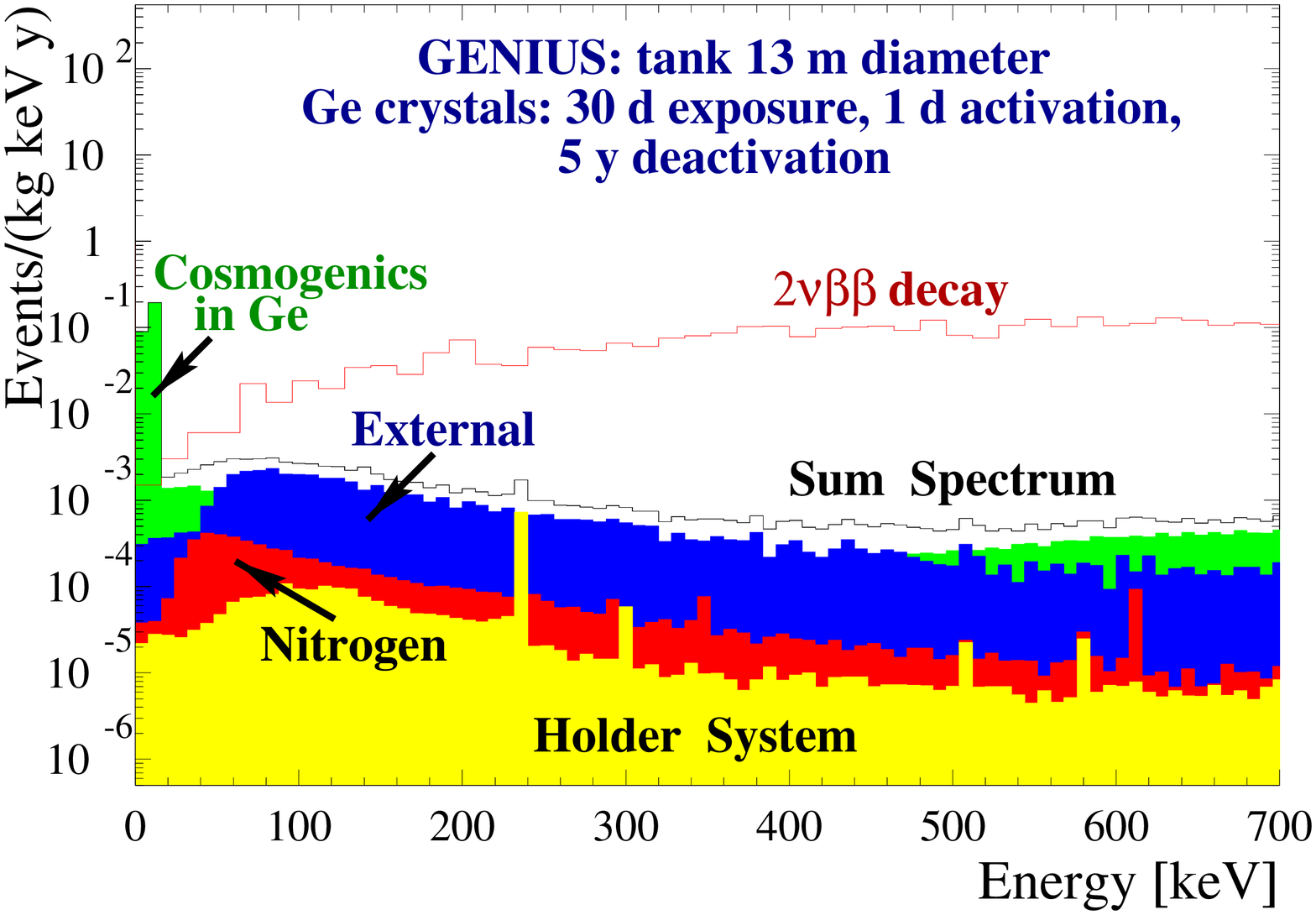}}
\caption{
       {\it Total background in a 13 m liquid nitrogen tank for detectors 
       produced as described in Fig. \ref{Cosmo-1d-3y} 
	(tritium neglected) (see 
\cite{KK01,KK-LowNu2}).}
\label{DBB-Spectr-13m-1d-5y-nonsat}}
\end{figure}



\section{GENIUS and Low-Energy Solar Neutrinos}
	GALLEX and SAGE measure $pp +\, ^7{\rm Be} +\, ^8{\rm B}$ 
	neutrinos (60 + 30 + 10\%) down to 0.24 MeV, 
	the Chlorine experiment measured $^7{\rm Be} +\, ^8{\rm B}$ 
		neutrinos (80\% $^8$B) above $E_\nu= 0.817$~MeV, 
		all without spectral, time and direction information. 
		No experiment has separately measured the 
		$pp$ and $^7$Be neutrinos and no experiment has 
		measured the {\em full}\ $pp$ $\nu$ flux. 
		BOREXINO plans to measure $^7$Be neutrinos, 
		the access to $pp$ neutrinos being limited by $^{14}$C 
		contamination (the usual problem of organic scintillators). 
		GENIUS could be the first detector measuring 
		the {\em full}\ $pp$ (and $^7$Be) 
		neutrino flux in real time.

		Extending the radius of GENIUS to 13 m and improving 
		some of the shielding parameters as described in 
\cite{BKK-SolN,GEN-prop} 
	the background can be reduced to a level of 
		$10^{-3}$ events/(kg y keV) (see also 
\cite{KKLowNu2}). 
Figure~\ref{Cosmo-1d-3y} 
	shows the simulated background from the cosmogenics produced 
       during detector production, assuming 30 d of exposure to cosmic 
       rays of the material between mining and zone refining, 1 d of 
       exposure during and after zone refining, 
       and 3 years of deactivation of the detectors in underground. 
Figure~\ref{DBB-Spectr-13m-1d-5y-nonsat} 
	shows the total background expected under these production 
       conditions.

\begin{figure}
\centering{
\includegraphics*[scale=0.5, angle=-90]
{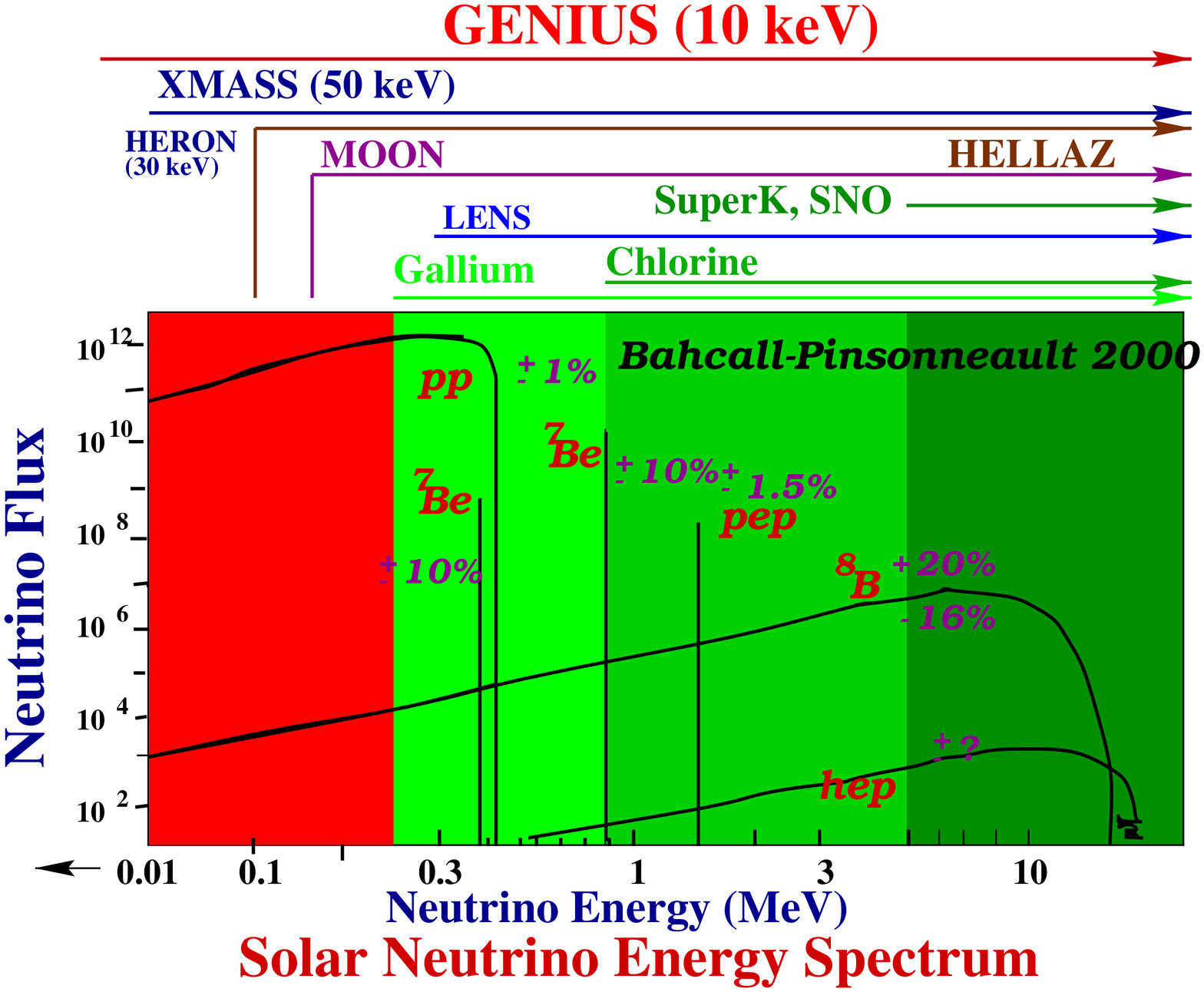}}
\caption[]{
       {\it The sensitivity (thresholds) of different running and projected 
       solar neutrino detectors (see 
\cite{HomP-Bach} and HEIDELBERG NON-ACCELERATOR PARTICLE PHYSICS GROUP 
	home-page: http://www.mpi-hd.mpg.de/non\_acc/).}
\label{sol-neutr-Bach}}

\centering{
\includegraphics*[scale=0.45, angle=-90]
{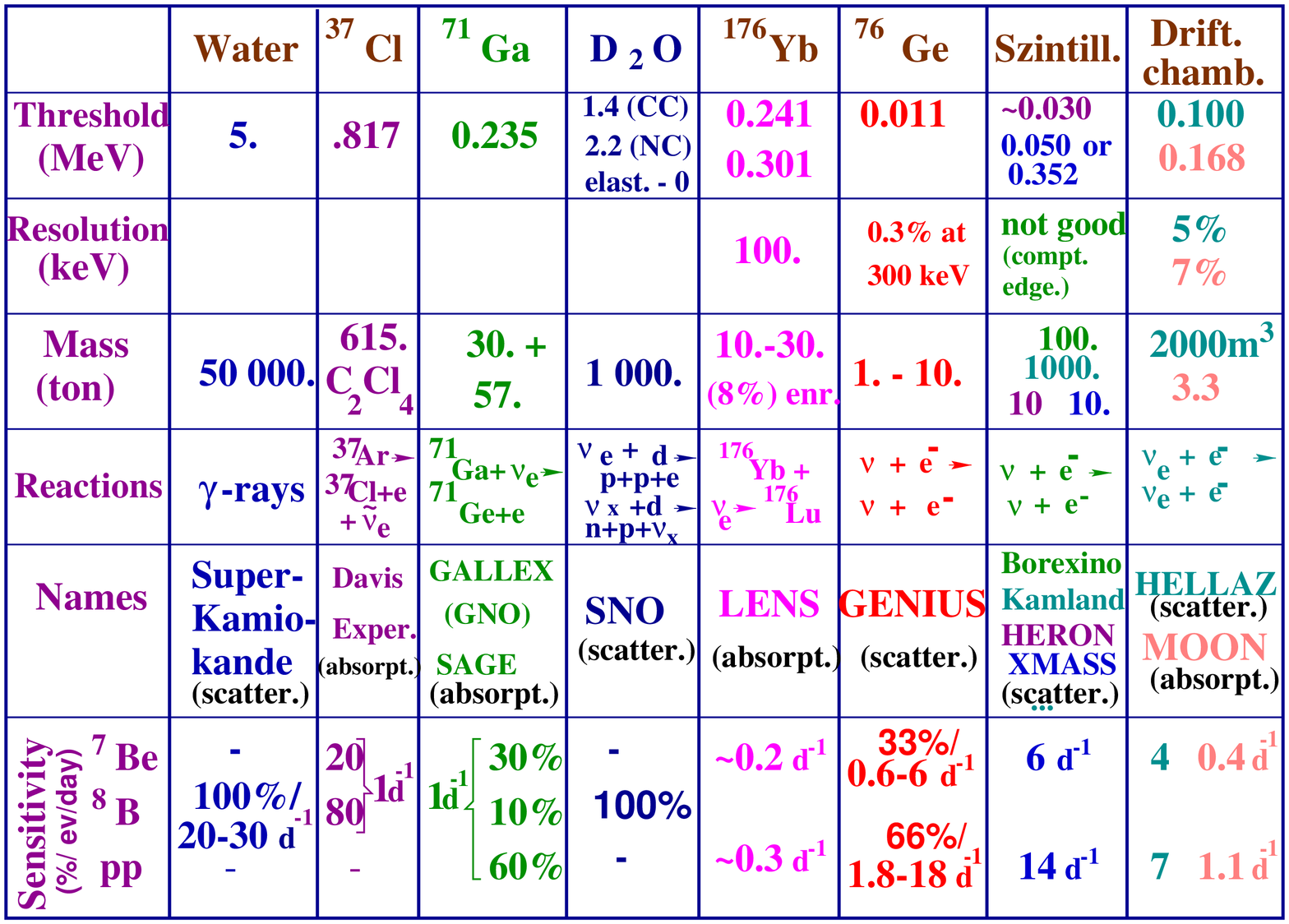}}
\caption{
       {\it Some key numbers of running and future solar neutrino 
       experiments (see also 
\cite{KK-LowNu2}).}
\label{tabl-solNeu-exper}}
\end{figure}

       This background will allow to look for the $pp$ and $^7$Be 
       solar neutrinos by elastic neutrino-electron scattering with a 
       threshold of 11 keV or at most 19 keV 
       (limit of possible tritium background)
(Fig.~\ref{sol-neutr-Bach},~\ref{pp_7be}) which would be the lowest threshold among other 
      proposals to detect $pp$ neutrinos, such as 
      HERON \cite{LowNu2}, 
HELLAZ \cite{LowNu2}, 
NEON \cite{LowNu2}, 
LENS \cite{Fuj00,LowNu2}, 
MOON \cite{Ej00,LowNu2}, 
XMASS \cite{XMAXX00,LowNu2}.

	The counting rate of GENIUS (10 ton) would be 6 events per day 
	for $pp$ and 18 per day for $^7$Be neutrinos, i.e. similar to 
	BOREXINO, but by a factor of 30 to 60 larger than a 20 ton LENS 
	detector and a factor of 10 larger than the MOON detector (see 
Fig.~\ref{tabl-solNeu-exper}).



\section{GENIUS-Test Facility}
		     Construction of a test facility for 
		     GENIUS --- GENIUS-TF --- 
		     consisting of $\sim$~40 kg of HP Ge 
		     detectors suspended in a liquid nitrogen 
		     box has been started. 
		     Up to end of January 2001, four detectors each 
		     of $\sim$~2.5 kg and with a threshold of as low 
		     as $\sim$~500 eV have been produced.

\begin{figure}
\centering{
\includegraphics*[scale=0.4]
{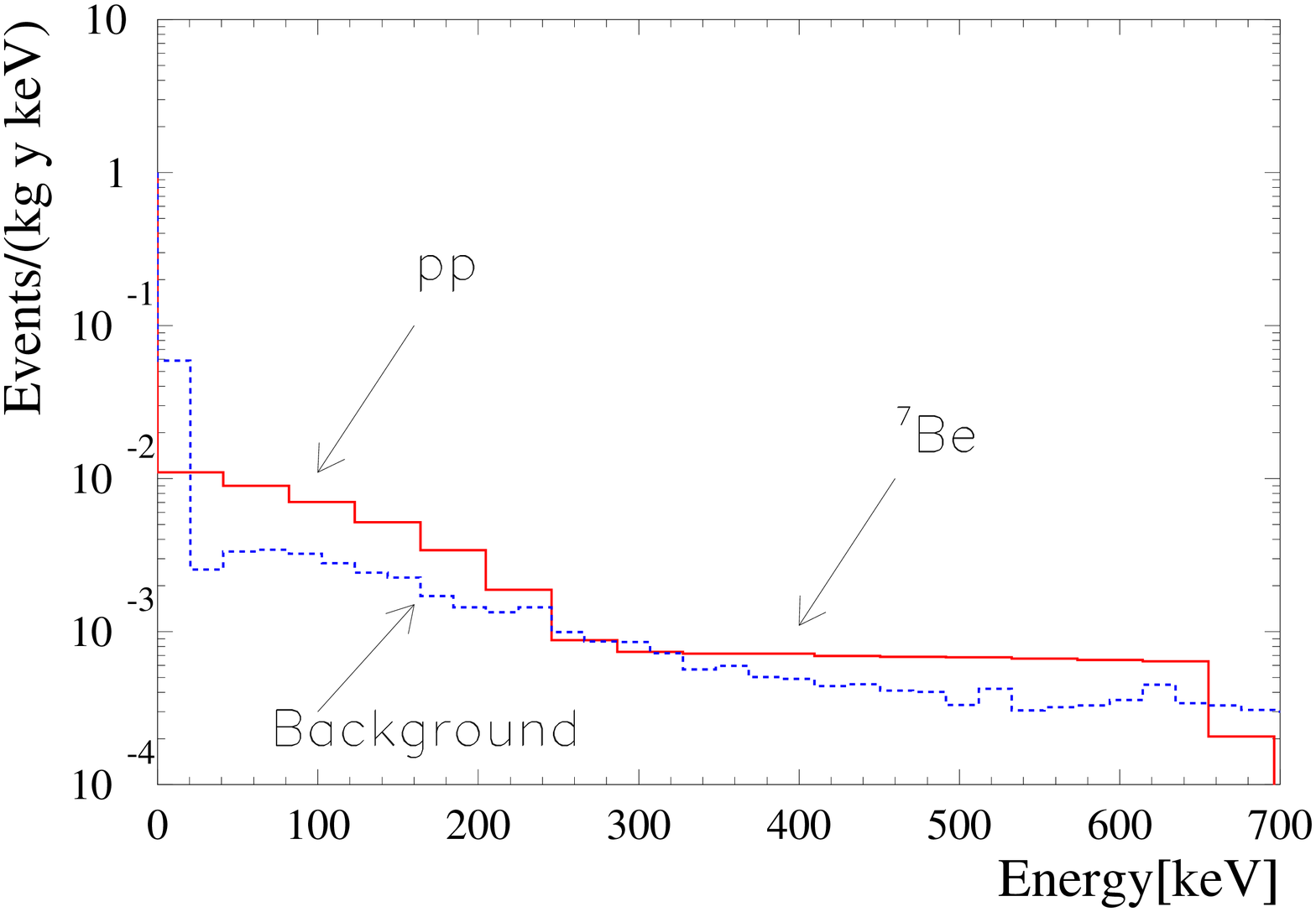}}

\vspace{-.5cm}
\caption[]{
       {\it Simulated spectrum of low-energy solar neutrinos 
       (according to SSM) for the GENIUS detector 
       (1 tonne of natural or enriched Ge) (from 
\cite{BKK-SolN}).}
\label{pp_7be}}
%
\centering{
\includegraphics*[scale=0.45]
{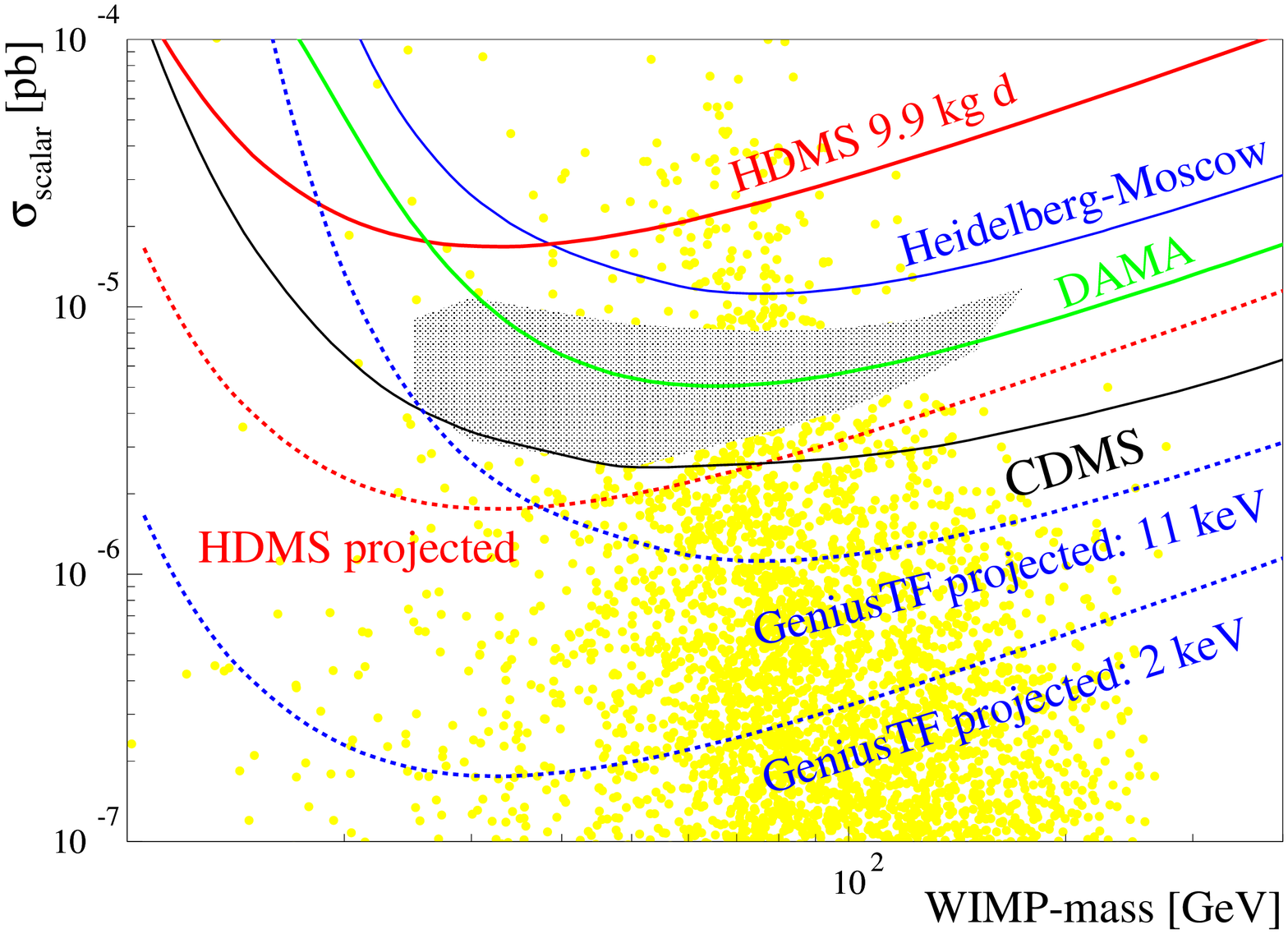}}

\vspace{-.5cm}
\caption[]{
       {\it WIMP-nucleon cross section limits as a function of the WIMP 
	mass for spin-independent interactions. 
       The solid lines are current limits of the HEIDELBERG-MOSCOW 
       experiment 
\cite{HM98}, the HDMS prototype 
\cite{KK-PRD63-00}, the DAMA experiment 
\cite{DAMA98}  
       and the CDMS experiment 
\cite{CDMS00}. 
       The dashed curves are the expectation for HDMS 
\cite{KK-PRD63-00} and for 
       GENIUS-TF with an energy threshold of 11 keV and 2 keV respectively, 
       and a background index of 2 events/(kg y keV) below 
       50 keV. 
       The filled contour represents the 2$\sigma$ evidence region 
       of the DAMA experiment 
\cite{DAMA98-99}. 
       The experimental limits are compared to expectations (scatter plot) 
       for WIMP-neutralinos calculated in the MSSM parameter space 
       under the assumption that all superpartner masses are lower than 
       300--400~GeV 
\cite{BedKK-01} (from 
\cite{KK-GeTF-MPI,GenTF-0012022}).}
\label{limits_gtf}}
\end{figure}

\begin{figure}
\centering{
\includegraphics*[scale=0.5]
{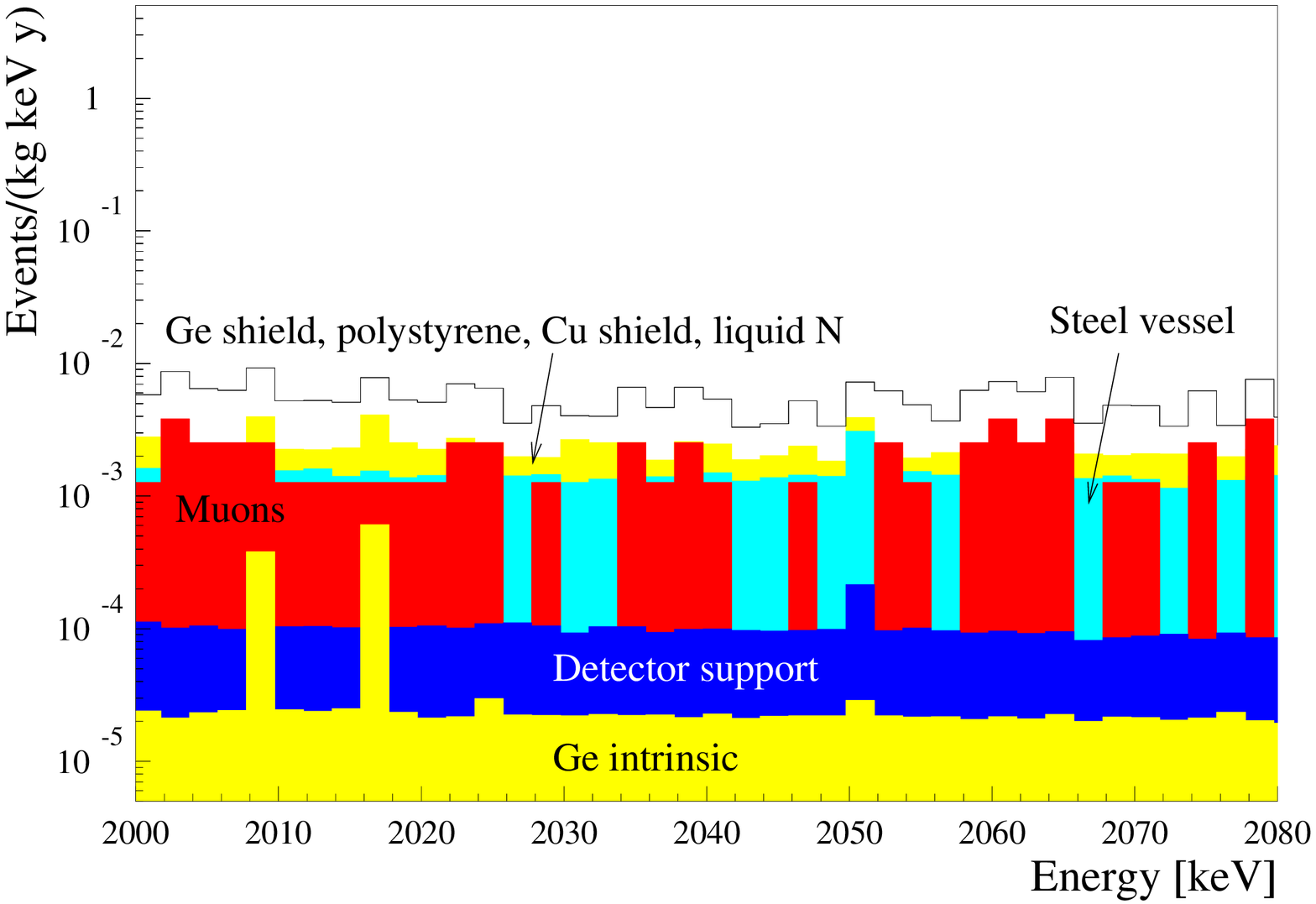}}
\caption[]{ 
       {\it Simulated spectra of the dominant background sources for the 
       enriched $^{76}$Ge detectors of the HEIDELBERG-MOSCOW experiment 
       in the GENIUS-TF setup. 
       The energy region relevant for the search of the neutrinoless 
       double beta decay is shown. 
       The solid line represents the sum 
       spectrum of all the simulated components (from 
\cite{KK-GeTF-MPI,GenTF-0012022}.)}
\label{sumspec_betabeta}}
%
\vspace{9pt}
\centering{
\includegraphics*[scale=0.5, angle=-90]
{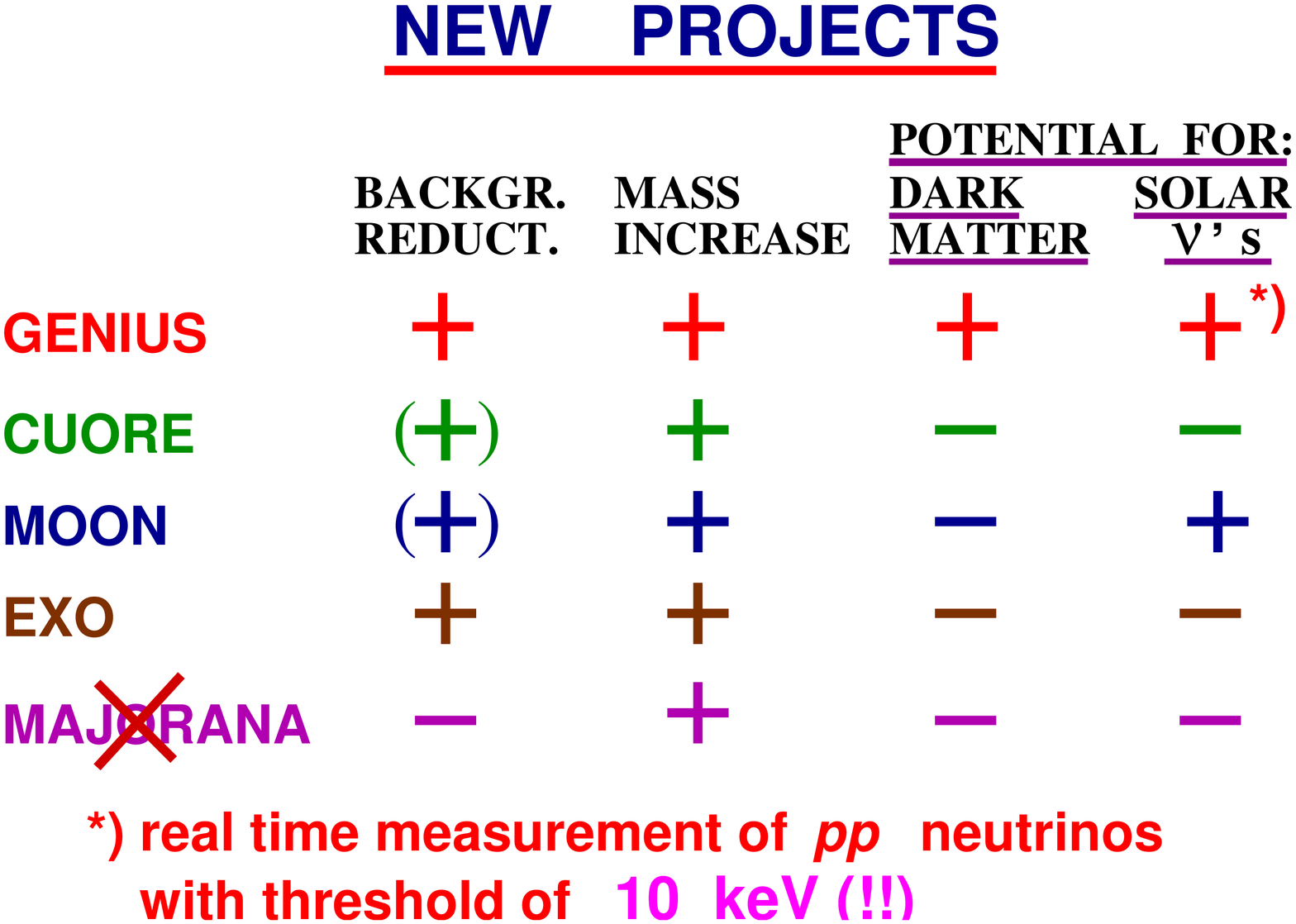}}
\caption[]{
       {\it New projects under discussion for future double beta 
       decay experiments (see 
\cite{KK60Y}).}
\label{NewProj1-catMaj}}
\end{figure}

         Besides test of various parameters of the GENIUS project, 
	 the test facility would allow, with the projected background 
	 of 2--4 events/(kg y keV) in the low-energy range, to probe the DAMA 
	 evidence for dark matter by the seasonal modulation signature 
	 within about one year of measurement with 95\% C.L.
	 Even for an initial lower mass of 20 kg the time scale would be 
	 not larger than three years, see 
Fig.~\ref{limits_gtf} (for details see 
\cite{KK-GeTF-MPI,GenTF-0012022}). 
	 If using the enriched $^{76}$Ge detectors of the HEIDELBERG-MOSCOW 
	 experiment in the GENIUS-TF setup, a background in the 
	 $0\nu\beta\beta$ region a factor 30 smaller than in the 
	 HEIDELBERG-MOSCOW experiment could be obtained, 
	 which would allow to test the effective Majorana neutrino mass 
	 down to 0.15 eV (90\% C.L.) in 6 years of measurement 
(Fig.~\ref{sumspec_betabeta}). 
      This limit is similar to what much larger experiments aim at 
(Table~1).



\begin{table*}
\caption{{\it Some key numbers of future double beta decay experiments (and of 
	the {\sf HEIDELBERG-MOSCOW} experiment). Explanations: 
	${\nabla}$ - assuming the background of the present pilot project. 
	$\ast\ast$ - with matrix element from  
	\cite{StMutKK90}, \cite{Tom91}, \cite{Hax84}, \cite{WuStKKChTs91}, 
	\cite{WuStKuKK92} (see Table II 
	in \cite{HM99}). 
	${\triangle}$ - this case shown 
	to demonstrate {\bf the ultimate limit} of such experiments. 
	For details see \cite{KK60Y}.}}
\label{table:1}
\newcommand{\m}{\hphantom{$-$}}
\newcommand{\cc}[1]{\multicolumn{1}{c}{#1}}
\renewcommand{\tabcolsep}{.65 pc} 
\renewcommand{\arraystretch}{.95} 
{\footnotesize
{  
\begin{tabular}[!h]{|c|c|c|c|c|c|c|c|}
\hline
\hline
 &  &  &  & Assumed &  &  & \\
 &  &  &  & backgr. & $Run-$ & Results & \\
$\beta\beta$-- & & & Mass & $\dag$ events/ & $ning$ & limit for & 
${<}m_{\nu}{>}$ \\
$Isoto-$ & $Name$ & $Status$ & $(ton-$ & kg y keV, & Time  
& $0\nu\beta\beta$ & \\
pe & & & $nes)$ & $\ddag$ events/kg & (tonn. & half-life & ( eV )\\ 
& & & & y FWHM,  & years) & (years) & \\
& & & & $\ast$ events &  &  & \\
& & &  & /yFWHM &  &  & \\
\hline
\hline
 &  &  &  &  &  &  & \\
~${\bf ^{76}{Ge}}$ & {\bf HEIDEL-} & {\bf run-}  & 0.011 & $\dag$ 0.06 
& {\bf 37.24} & ${\bf 2.1\cdot{10}^{25}}$ & {\bf $<$ 0.34} $\ast\ast$\\
 & {\bf BERG}  & {\bf ning}  &  (enri-  &  &  {\bf kg y} &  {\bf 90$\%$ c.l.} 
& {\bf 90$\%$ c.l.} \\
& {\bf MOSCOW} & {\bf since} & ched) & $\ddag$ 0.24  &  
& ${\bf 3.5\cdot{10}^{25}}$ & {\bf $<$ 0.26} $\ast\ast$\\
& {\bf \cite{KK-SprTracts00}} & {\bf 1990} &  & $\ast$ 2 & & 
{\bf 68$\%$ c.l.} & {\bf 68$\%$ c.l.}\\
& {\bf \cite{AnnRepGrSs00,KK01}} &  &  &  &  & {\bf NOW !!}  & {\bf NOW !!}\\
\hline
\hline
\hline
 &  &  &  &  &  &  & \\
${\bf ^{100}{Mo}}$ & {\sf NEMO III} & {\it under} & $\sim$0.01 & $\dag$ 
{\bf 0.0005} &  &  &\\
 & {\tt \cite{NEMO-Neutr00}}& {\it constr.} & (enri- & $\ddag$ 0.2  & 50 & 
${10}^{24}$ & 0.3-0.7\\
 &  & {\it end 2001?} & -ched) &  $\ast$ 2 &kg y  &  &\\
\hline
\hline
&  &  &  &  &  &  & \\
${\bf ^{130}{Te}}$ & ${\sf CUORE}^{\nabla}$ & {\it idea} & 0.75 & $\dag$ 0.5 
& 5 & $9\cdot{10}^{24}$ & 0.2-0.5\\
 & {\tt \cite{CUORE-LeptBar98}}& {\it since 1998} &(natural)  
& $\ddag$ 4.5/$\ast$ 1000 &  & & \\ 
\hline
&  &  &  &  &  &  &  \\
${\bf ^{130}{Te}}$ & {\sf CUORE}  &  {\it idea} & 0.75   & $\dag$ 0.005 & 5 
& $9\cdot{10}^{25}$ & 0.07-0.2\\
&  {\tt \cite{CUORE-LeptBar98,Fior-Neutr00}} & {\it since 1998}   & (natural)  
&  $\ddag$ 0.045/ $\ast$ 45 &  & &\\
\hline
&  &  &  &  &  &  & \\
${\bf ^{100}{Mo}}$ & {\sf MOON} & {\it idea} & 10 (enrich.) & ? & 30 & ? &\\
 & {\tt \cite{Ej00,LowNu2}} & {\it since 1999} &  100(nat.) & & 300 & &0.03 \\
\hline
&  &  &  &  &  &  & \\
${\bf ^{116}{Cd}}$ & {\sf CAMEOII} & {\it idea}  & 0.65 & * 3. & 5-8  
& ${10}^{26}$ & 0.06 \\
& {\sf CAMEOIII}{\sf \cite{Bell00}} & {\it since 2000 } & 1(enr.) & ? & 5-8 
&  ${10}^{27}$ & 0.02 \\
\hline
&  &  &  &  &  &  & \\
${\bf ^{136}{Xe}}$ & {\sf EXO} & Proposal& 1 & $\ast$ 0.4 & 5 & 
$8.3\cdot{10}^{26}$ & 0.05-0.14\\
&  & since &  &  &  &  & \\
  & {\tt \cite{EXO00,EXO-LowNu2}} & 1999 & 10 & $\ast$ 0.6 & 10 & 
$1.3\cdot{10}^{28}$ & 0.01-0.04\\
\hline 
\hline
\hline
\hline
&  &  &  &  &  &  &  \\
~${\bf ^{76}{Ge}}$ & {\bf GENIUS} & {\it under} & 11 kg & 
$\dag$ ${\bf 6\cdot{10}^{-3}}$& 3 
& {\bf ${\bf 1.6\cdot{10}^{26}}$} & {\bf 0.15} \\
& {\bf - TF} & {\it constr.} & (enr.) &  &  &  &   \\
&  {\bf \cite{KK-GeTF-MPI,GenTF-0012022}}&  {\it end 2001?} &  & &  &  &  \\
\hline
&  &  &  &  &  &  &  \\
~${\bf ^{76}{Ge}}$ & {\bf GENIUS} & Pro- & 1  & $\dag$ 
${\bf 0.04\cdot{10}^{-3}}$ & 1 & ${\bf 5.8\cdot{10}^{27}}$ & 
{\bf 0.02-0.05} \\
 & {\tt \cite{KK-Bey97,GEN-prop}}  & posal &(enrich.)  
& $\ddag$ ${\bf 0.15\cdot{10}^{-3}}$ & & & \\
&  & since &  & $\ast$ {\bf 0.15} &  &  &  \\
&  & 1997 & 1 & ${\bf \ast~ 1.5}$ & 10 & ${\bf 2\cdot{10}^{28}}$  & 
{\bf 0.01-0.028} \\
\hline
&  &  &  &  &  &  &  \\
~${\bf ^{76}{Ge}}$ & {\bf GENIUS} & Pro- & 10 
& $\ddag$ ${\bf 0.15\cdot{10}^{-3}}$ & 10 &
${\bf 6\cdot{10}^{28}}$ & {\bf 0.006 -}\\
&  {\tt \cite{KK-Bey97,GEN-prop}} &  posal &  &  &  &  &  {\bf 0.016}\\
 &   &  since &(enrich.) & ${\bf 0^{\triangle}}$ & 10 & 
${\bf 5.7\cdot{10}^{29}}$ & {\bf 0.002 -}\\
&  &  1997 &  &  &  &  &  {\bf 0.0056}\\ 
\hline 
\hline
\end{tabular}\\[2pt]
}}
\end{table*}


\section{Conclusion}
	The GENIUS project is --- among the projected or discussed other 
	third generation double beta detectors --- the one which exploits 
	this method to obtain information on the neutrino mass to the 
	ultimate limit. 
	Nature is extremely generous to us, that with an increase of 
	the sensitivity by two orders of magnitude compared to the 
	present limit, down to 
$\langle m_\nu\rangle < 10^{-3}$~eV, 
	 indeed essentially all neutrino scenarios allowed by present 
	 neutrino oscillation experiments can be probed.

	 GENIUS is the only of the new projects 
(Fig.~\ref{NewProj1-catMaj}) which simultaneously has a huge potential for 
      cold dark matter search, and for real-time detection of 
      low-energy neutrinos.


\end{document}